# Microwave Quenching in DC-Biased Coplanar Waveguide Based on YBa$_2$Cu$_3$O$_{7-\delta}$ Thin Film

Nickolay T. Cherpak, *Senior Member, IEEE*, Alexey I. Gubin, Alexander A. Lavrinovich, and Svetlana A. Vitusevich

*Abstract*—In this paper, we report on the results of features in the discovered effect of a strong change in the microwave losses (quenching) in HTS-based coplanar waveguide (CPW) at certain values of the input power $P_{in}$ and direct current (dc) $I_{dc}$. Two waveguides were studied: CPW-150 and CPW-75, fabricated of epitaxial 150- and 75-nm YBCO films. The insertion loss of CPW IL= 10 lg ($P_{out}/P_{in}$) was measured versus both the operating temperature and the bias dc current $I_{dc}$ when a fixed level of the input X-band pulsed MW signal $P_{in}$ was applied. Here $P_{out}$ is the power measured at the CPW output. A pulse duration $\tau_i$ was 5 µs with pulse repetition period T = 40 µs. Experiments showed a noticeable difference in specific values of $I^2 = I_{dc}$ corresponding to the $P_{in}$-dependent quenching phenomenon in CPW-150 and CPW-75. With a weak input signal, the difference in behavior of CPWs is anticipated, i.e., quenching occurs at a lower current for thinner film. For $P_{in}$ > 1 W the situation is opposite, i.e., quenching for CPW-150 takes place at a lower $I^2$. This is explained by stronger microwave heating of CPW-150.

## I. Introduction

SOON after the discovery of high-temperature superconductivity (HTS) [1] the idea of creating MW power limiter using surface resistance $R_s$ dependenc eon input power $P_{in}$ was suggested and some of the first experiments were performed with a planar transmission line [2], [3]. This approach was originally developed in [4], [5]. Despite the possibility of constructing such a limiter, serious obstacles were found in the implementation of the concept into practical devices: the difficulty to control the input power handling $P_{in}$ [4] and the instability of planar transmission line upon the transition to a strongly dissipative condition [5]. In [6] it was proposed to control the nonlinear impedance of a coplanar HTS thin film waveguide (CPW) by using dc bias current.

Recently we reported on the effect of a strong change in the microwave losses in HTS-based CPW at certain values of the input power $P_{in}$ and direct current (dc) $I_{dc}$ [7]. A sharp transition of HTS structure into a strongly dissipative state was observed when passing dc through the structure. The 75 and 150 nm thin YBa$_2$Cu$_3$O$_{7-\delta}$ (YBCO) films CPW on a single crystal MgO substrate was studied experimentally. A sharp and reversible transition of the CPW into a strongly dissipative state was observed at certain values of $P_{in}$ and $I_{dc}$ that depended on temperature. In this paper, we report on the results of studying the new peculiarities in the discovered effect.

TABLE I
CHARACTERISTICS OF YBCO EPITAXIAL FILMS ON THE SINGLE CRYSTAL MgO SUBSTRATES

| Film thickness (nm) | Critical temperature (K) | Critical current density at 77K (MA/cm$^2$) |
|---|---|---|
| 150 | 86.5 | 3.6 |
| 75 | 85.8 | 3.0 |

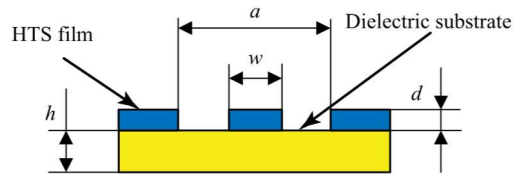

Fig. 1. Cross section of YBCO-film-based coplanar waveguide.

## II. Experimental Details

The quenching phenomenon was studied in two CPWs: CPW-150 and CPW-75, fabricated of epitaxial 150 nm and 75 nm thick YBCO epitaxial films. The HTS films epitaxially grown by THEVA (Germany) have the characteristics listed in Table I. Microwave studies showed their very good microwave impedance properties [8].

CPWs were patterned using photolithography. The crosssection dimensions of the structures are $a$ =0.186 mm, $w$ = 0.1 mm, and $h$ = 0.5 mm (Fig. 1). The waveguides are represented by 16.8 mm long straight sections with gold-plated pads, through which the feed-in and read-out of the microwave signal and direct current (dc) transmission were simultaneously measured by using integrated planar bias tees (IPBT). Continuous signal from a magnetron source operated at $f$ = 9.24 GHz, with the output power adjustable in the range of 0 to 13 watts was modulatedby the square-wavepulse generator with the following characteristics: pulse duration $\tau_i$ = 5 µs, the pulse repetition period $T$ =40 µs (repetition frequency $f_r$ = 2.5×10$^4$ Hz). The measured CPW was positioned inside a chamber filled with helium gas and cooled by liquid nitrogen. The insertion loss of CPW, $IL$ = 10 lg ($P_{out}/P_{in}$), was studied versus both temperature of the cooled chamber and the bias dc current $I_{dc}$ when a fixed level of the input X-band pulsed MW signal $P_{in}$ was applied.

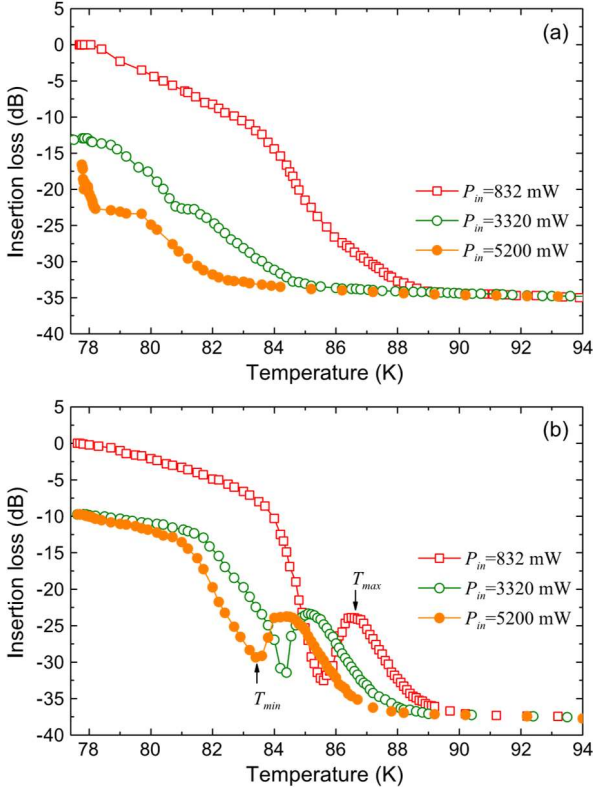

Fig. 2. Insertion loss in (a) CPW-150 and (b) CPW-75 depending on temperature at different input power of 5-µs-long MW pulses.

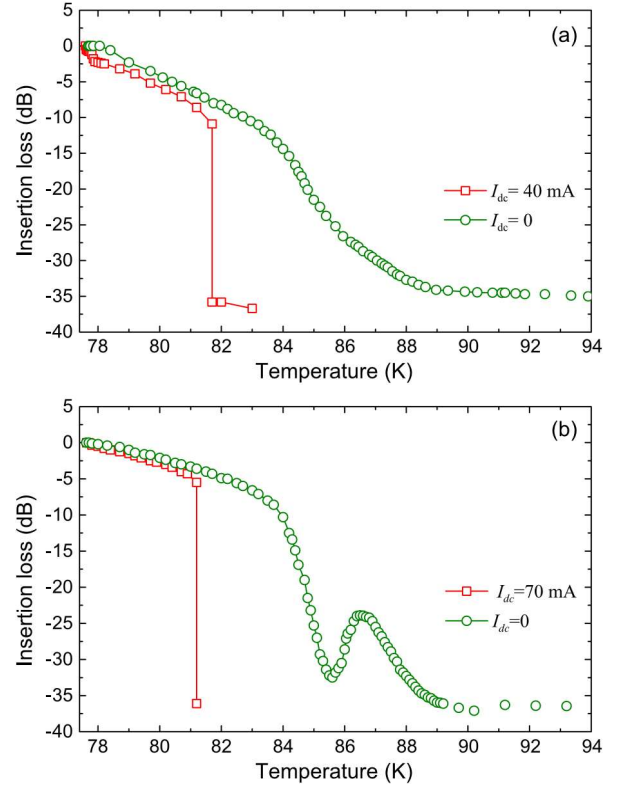

Fig. 3. Insertion loss in (a) CPW-150 and (b) CPW-75 depending on temperature at 5 µs-long pulse of $P_{in}$ = 832 mW at dc ($I_{dc}$ = 40 mA for CPW-150 and 70 mA for CPW-75) and without it ($I_{dc}$ = 0).

III. RESULTS AND DISCUSSION

Fig. 2 shows the insertion loss dependence on temperature $T$ for several values of $P_{in}$: (a) for CPW-150 and (b) for CPW-75nm. One can see that at a small $P_{in}$ the plot $IL(T)$ is monotonic (Fig. 2(a)). However, with increasing $P_{in}$, along with a general increase in CPW-150 losses, the nonmonotonic steps appear in $IL(T)$. The overall increase in losses is caused by an increase of the surface resistance $R_s(H_\omega)$ or $R_s(P_\omega)$ [9] ($H_\omega$ and $P_\omega$ are the MW magnetic field amplitude and the power in the CPW). The appearance of these steps can be explained by the formation of thermal domains [10], [11] and/or the wave reflection from the input and output of the CPW [12].

The dependence $IL(T)$ is in CPW-75 is displayed in Fig. 2(b) for several values of $P_{in}$. Two extremes: minimum and maximum are revealed in each of the curves instead of steps, observed in the case of for CPW-150 (see Fig. 2(a)). The increase in power leads to a shift of the "maximum pump hump" to the lower temperatures and expanding its top parts.

This dependence changes dramatically when direct current (dc) $I_{dc}$ is passed through the CPW-150 (Fig. 3(a)). It's seen that if the values of $P_{in}$ and $I_{dc}$ are fixed, then at a low temperature there is a sharp transition of the CPW into a strongly dissipative state at a certain temperature. The jump in $IL$ is recorded to be of almost three orders of magnitude in the particular case for $I_{dc}$ = 40 mA [7]. Strong switching effect is also revealed for CPW-75 (Fig. 3(b)).

Avalanche growth of the microwave loss $IL$, i.e., microwave quenching stimulated by dc, depending on operating temperature is shown in Fig. 4. The effect can be seen on the $IL(I_{dc})$ curves at $T$ = const and different values of $P_{in}$ (Fig. 5) and on the $IL(P_{in})$ curves at $I_{dc}$ = const (not shown in the figure). The quenching phenomenon takes place at a certain value of $I_{dc} = I^*$. Fig. 5 shows that $I^*$ value decreases with increasing $P_{in}$. Thus for the quenching process both current components, microwave $I_\omega$ and dc $I_{dc}$ currents, play a crucial role.

The ratio of dc and MW current densities, $J_{dc}^*$ and $J_\omega$, triggering the quenching effect, was found in [7] as a function of the input power $P_{in}$ for CPW-150. It was shown, the maximum value of the sum of the current densities $J_{dc}$ and $J_\omega$ always remains below the value of the threshold density of the critical current $J_c$ = 3.6×10$^6$ A/cm$^2$ for CPW-150. This fact can be understood by assuming additivity of the effects at direct and microwave currents in the transmission line. At the same time, if we assume (for simplicity) homogeneous current distribution in the cross section of the central strip of the CPW, it would be expected that the sum $J_{dc}^* + J_\omega$ is independent of $P_{in}$. However, the sum $J_{dc}^* + J_\omega$ varies depending on the input power, while remaining less than the critical current density (Fig. 6). The important role of the heat effects in the observed feature is confirmed by not only peculiarities, pointed in Fig. 6, but by a stronger response of the resistive transition in the CPW-150 compared with weak response in the CPW-75 (it is not shown here) as well. Using of the magnetic flux flow effects in the explanation of the observed feature is reasonable because the current density at the edges of the central CPW nar waveguides. (a) CPW-150 Fig. 4. Microwave quenching in dc-biased YBa($I_{dc}$ = 40 mA2)Cu and (b) CPW-75 3O7−δ film-based copla- ($I_{dc}$ = 70 mA) depending on temperature at certain $I_{dc}$ and different $P_{in}$.

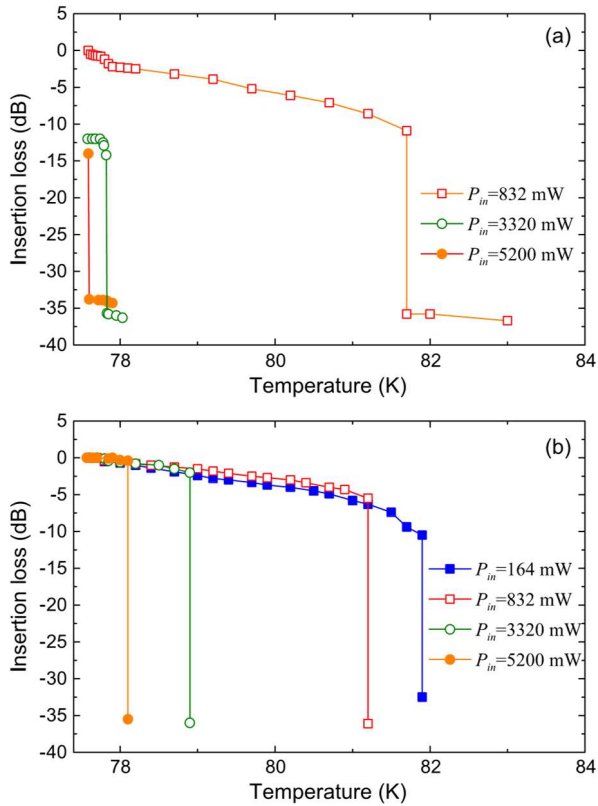

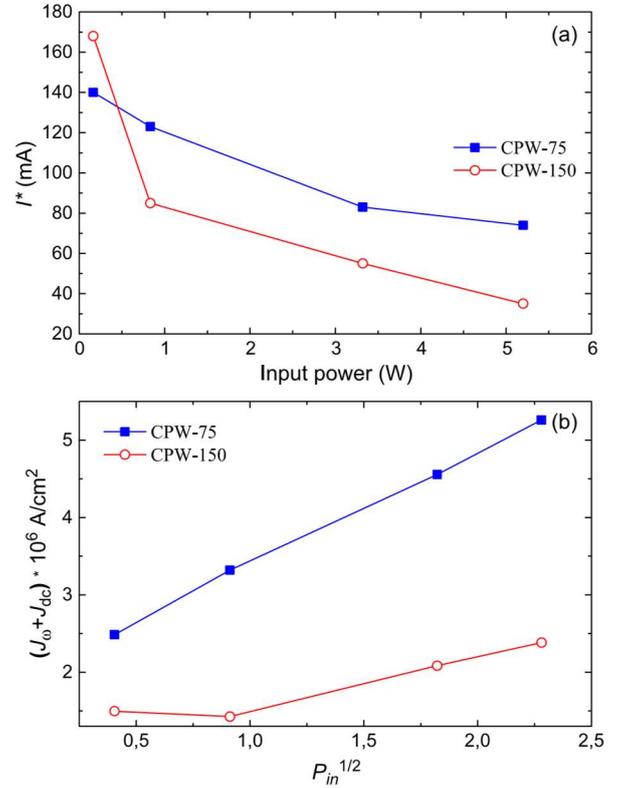

Fig. 6. Direct current values corresponding to the quenching phenomenon $I*$ (a) as functions of the input power $P_{in}$ and (b) sum of current densities $J^*_{dc} + J_\omega$ as function of the square root of the input power $P_{in}$ in CPW-150 and CPW-75 at $T$ = 77 K.

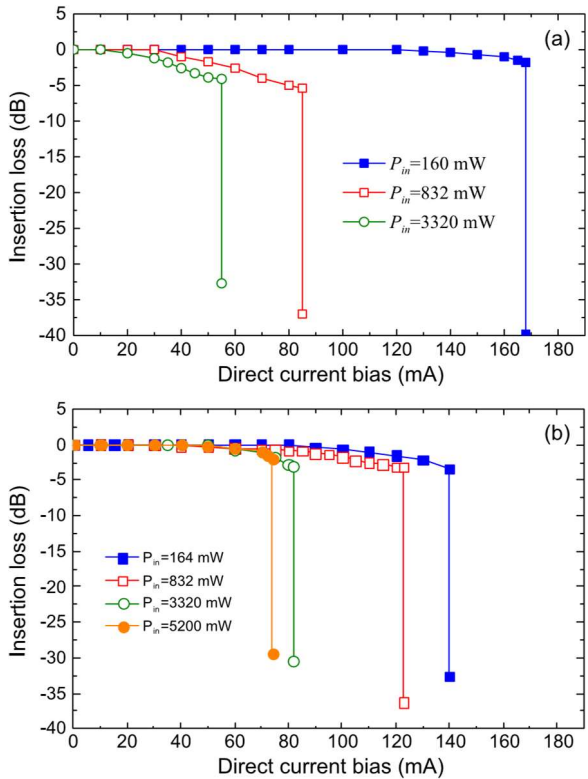

Fig. 5. Insertion loss in (a) CPW-150 and (b) CPW-75 depending on direct current at different values of the input power $P_{in}$ at $T$ =77 K.

line is always higher the critical current density as the result of inhomogeneous current redistribution in superconducting structure [13]. However, we do not know the exact physical phenomenon of the additional strong nonlinear effect inclusion with passing the dc through the structure. In other words, the mechanism of the "a trigger hook" is unknown.

One of the possible explanations could be describing based on known peak-effect, which appears as in the case of dc [14], [15] as in the microwave current [16], [17]. At the same time, we should note, that our investigations were carried out in the absence of dc magnetic field in difference with [16], [17]. The authors hope that the nature of the observed effect will become clear in the course of further research, including study in dc magnetic field.

It is reasonable to compare behavior of both dc-biased CPWs as a function of $P_{in}$. Fig. 6(a) shows noticeable and unexpected to some extent difference in recorded dependences of a specific values of $I*$ = $I_{dc}$ corresponding to the quenching phenomenon as a function of $P_{in}$ for CPW-150 and CPW-75.

At weak input signal the difference in behavior of CPWs is anticipated, i.e., quenching occurs at a lower current for thinner film. However, for $P_{in}$ > 1 W situation is opposite, the quenching for CPW-150 takes place at a lower $I*$.

This behavior can be explained by stronger microwave heating of CPW-150. We can suppose, the transition occurs at lower powers in CPW-150 because the heat accumulated during a pulse in CPW-150 per unit of the film-substrate interface is larger than that accumulated in CPW-75 while the thermal resistance of the interface is the same [18]. This explains why the sum of current densities $J_{dc}^* + J_\omega$ as a function of $P_{in}$, which determines the quenching effect, is lower for the thicker film structure, namely for CPW-150 (see Fig. 6(b)).

IV. CONCLUSION

In this paper, we report on the behavior of the earlier discovered effect of microwave quenching in dc-biased HTS coplanar waveguide (CPW). This phenomenon was studied in two CPWs fabricated of epitaxial 150 nm and 75 nm YBCO films. In the experiments, the insertion loss of CPW was measured versus both temperature of the cooled chamber and the bias dc current $I_{dc}$ when a fixed level of the input X-band pulsed MW signal $P_{in}$ was applied. Experiments showed a noticeable and in some extent unexpected difference in the observed dependence of the specific value of $I^* = I_{dc}$ corresponding to the quenching phenomenon as a function of $P_{in}$ for the waveguides. With a weak input signal the difference in behavior of CPWs is anticipated, i.e., quenching occurs at a lower current for thinner film. However, for $P_{in} > 1$ W situation is opposite, the quenching for CPW-150 takes place at a lower $I^*$. This is explained by stronger microwave heating of CPW-150.

REFERENCES

[1] J. G. Bednorz, and K. A. Müller, "Possible high T$_c$ superconductivity in the Ba-La-Cu- O system", *Z. Phys.-Condens. Matter* B, vol. 64, pp. 189-193, Apr. 1986.
[2] M.M.Gaidukov, S.G.Kolesov, L.Kowalewicz, A.B.Kozyrev, and O.G.Vendik, "Microwave power limiter based on high-Tc superconductive film", *Electron. Lett*. vol. 26, no. 16, pp. 1229-1231, Aug. 1990.
[3] V.N.Keis, A.B.Kozyrev, T.B.Samoilova, and O.G.Vendik, "High speed microwave filter-limiter based on high-T$_c$ superconducting films", *Electron. Lett.* vol. 29, no. 6, pp.546-547, Mar. 1993.
[4] J. C. Booth, D. A. Rudman, and R. H. Ono, "A Self-Attenuating Superconducting Transmission Line for Use as a Microwave Power Limiter", *IEEE Trans. Appl. Supercond.*, vol. 13, no. 2, pp. 305-310, Jun. 2003.
[5] A.A. Lavrinovich, E.V. Khramota, and N.T. Cherpak, "Investigation of the superconducting microwave transmission line in strong electromagnetic fields", *Telecomm. and Radio Engineer.*, vol. 68, no. 19, pp. 1741-1750, 2009.
[6] N.T.Cherpak, A.A.Lavrinovich, A.A.Kalenyuk, V.M.Pan, A.I.Gubin, E.V.Khramota, A.A.Kurakin, and S.A.Vitusevich, "DC-biased coplanar waveguide on the basis of high-T$_c$ superconducting thin film with nonlinear impedance", *Telecomm. and Radio Engineer.*,vol.69, no. 15, pp.13571364, 2010.
[7] N. T. Cherpak, A. A. Lavrinovich, A. I. Gubin, and S. A. Vitusevich, "Direct-current-assisted microwave quenching of YBa$_2$Cu$_3$O$_{7-\delta}$ coplanar waveguide to a highly dissipative state ", *Appl. Phys. Lett.*vol. 105, pp. 022601-1-022601-3, Jul. 2014.
[8] Barannik, N. T. Cherpak, M. S. Kharchenko, R. Semerad, and S. A. Vitusevich, "Surface impedance of YBa$_2$Cu$_3$O$_{7-\delta}$ films grown on MgO substrate as a function of film thickness", *J. Supercond. Novel Magn.*, vol.26, no. 4, pp.43-48, Dec. 2013.
[9] M. Hein, High-Temperature- Superconductor thin Films at Microwave Frequencies, Berlin-Heidelberg, Springer-Verlag, 1999.
[10] A.N. Resnik, A.A. Zharov, and M.D. Chernobrovtseva, "Nonlinear Thermal Effects in the HTSC Microwave Stripline Resonator", *IEEE Trans. Appl. Supercond.*, vol.5, no.2, pp. 2579-2582, Jun. 1995.
[11] A.A. Kalenyuk, "Nonlinear microwave response of superconducting YBa2Cu3O7–δ transmission strip line with constriction", *Low Temp. Phys.*, vol. 35, no. 2., pp.105-111, Feb. 2009.
[12] M. Ruibal, G. Ferro, M. R. Osorio, J. Maza, J. A. Veira, and F. Vidal, "Size effects on the quenching to the normal state of YBa$_2$Cu$_3$O$_{7-\delta}$ thin-film superconductors", *Phys. Rev. B*, vol. 75, pp. 012504-1-012504-4, Jan. 2007.
[13] A.Porch, M.J.Lancaster and R. Humphreys "Coplanar resonator technique for the determination of the surface impedance of patterned thin films", *IEEE Trans. on Microwave Theory and Techniques,* vol. **43,** no.2, 306-314, 1995.
[14] A.I. Larkin, and Yu.N. Ovchinnikov, "Pinning in type II superconductors", *J. Low Temp. Phys.*, vol. 34, no. 3, pp. 409-428, Feb. 1979.
[15] A.I. Larkin, M.C. Marchetti, and V.M. Vinokur, "Peak Effect in Twinned Superconductors", *Phys. Rev. Lett.***75**, 2992 (1995)
[16] A. R. Bhangale, P. Raychaudhuri, T. Banerjee, R. Pinto, and V. S. Shirodkar, "Peak effect in laser ablated DyBa$_2$Cu$_3$O$_{7-\delta}$ films at microwave frequencies at subcritical currents", *J. of Appl. Phys.*, vol. 89, pp. 7490-7492, Jul. 2001.
[17] T. Banerjee, D. Kanjilal, and R. Pinto, "Peak effect and its evolution with defect structure in YBa$_2$Cu$_3$O$_{7-\delta}$ thin films at microwave frequencies", *Phys. Rev. B*, vol. 65, pp. 174521, Jun. 2002.
[18] T. Banerjee, V. C. Bagwe, J. John, S. P. Pai, and R. Pinto, "Vortex dynamics at subcritical currents at microwave frequencies in DyBa$_2$Cu$_3$O$_{7-d}$ thin films," *Phys. Rev. B, Condens. Matter*, vol. 69, Mar. 2004, Art. no. 104533.